\documentclass[prc,twocolumn,showpacs,preprintnumbers,amsmath,amssymb]{revtex4-1}

\usepackage{amsmath,amssymb,multirow,epsfig,bm,color}

\newcommand{\beq}{\begin{equation}}
\newcommand{\eeq}{\end{equation}}
\newcommand{\bea}{\begin{eqnarray}}
\newcommand{\eea}{\end{eqnarray}}

\newcommand{\benn}{\begin{displaymath}}
\newcommand{\eenn}{\end{displaymath}}
\newcommand{\angstr}{\rm\AA}
\newcommand{\lin}{$2\mbox{D}\mbox{Li}_2\mbox{N}$}
\newcommand{\linh}{2DLi$_{2}$N+H}
\newcommand{\linhh}{2DLi$_{2}$N+2H}
\newcommand{\linli}{2DLi$_{2}$N+Li}
\newcommand{\linlili}{2DLi$_{2}$N+2Li}

\begin{document}

\title{Electronic and magnetic properties of two-dimensional Li$_{3}$N}

\author{K. Zberecki$^1$, M. Wierzbicki$^1$}
\affiliation{$^1$Faculty of Physics, Warsaw University of Technology, ul. Koszykowa 75, 00-662 Warsaw, Poland}
\email{zberecki@if.pw.edu.pl}
\date{\today}

\begin{abstract}
Using first-principles plane-wave calculations study of electronic and magnetic properties of hypothetical two-dimensional structure
of Li$_{2}$N compound have been conducted. Calculations show, that electronic properties of this this structure can be inflenced
by hydrogenation, which may change the system from wide-gap semiconductor to metal. Also, non-zero magnetic moment, equal to 1 $\mu_{B}$ 
can be generated by intruduction of H vacanies in hydrogenated structure.  
\end{abstract}


\maketitle

\section{INTRODUCTION}
Since its discovery in 2004 graphene ~\cite{nov1} draws much attention because of unique features of this 
two-dimensional system. Graphene is composed of a sp$^ {2}$-bonded carbon atoms forming honeycomb structure. 
It became famous for its very interesting electronic structure with characteristic, linear energy dispersion 
near K point of Brillouin zone and many other features ~\cite{graph1}. \newline
Shortly after, experimental techniques allowed fabrication of other new two-dimensional materials,
like BN and MoS$_2$ honeycomb structures ~\cite{nov2}. The discovery of such stable two-dimensional 
materials triggered search for similar structures made from different compounds. Up to now many of these 
hypothetical structures constructed from silanene (2D Si) and germanene (2D Ge) ~\cite{sil1, sil2}, 
III-V compounds ~\cite{III-V1}, SiC ~\cite{sic1} or ZnO ~\cite{zno2} have been studied theoretically. 
Also, calculations show ~\cite{bor1}, that graphene-like type of structure is not the only one possible 
for two-dimensional material. This new class of boron sheets, composed of triangular and hexagonal motifs 
can be stabilized by interplay of three- and two-center bonding scheme ~\cite{bor2}. Another example of 
triangular sheet could be found in already known material, which is Li$_{3}$N in its $\alpha$ phase. \newline 
Li$_3$N is a bulk material known to be a fast ion conductor ~\cite{brak}. Li$_3$N is also known as a candidate 
for hydrogen storage material due its high theoretical H$_{2}$ capacity ~\cite{lin3}. Bulk Li$_{3}$N crystallizes 
in hexagonal structure which is characterized by $P6/mmm$ symmetry group, each nitrogen atom is surrounded by eight
lithium atoms. It has layered structure, one layer is Li$_{2}$N and the other is of Li atoms only. Previous theoretical
studies confirm ionic nature of bonding in this compound ~\cite{lin1,lin2}. Since N-containing layer is rather weakly 
bound with two Li-only layers, it would be interesting to study electronic properties of such two-dimensional structure 
(2DLi$_{2}$N) - Fig \ref{fig0}a. Since this structure would have N atoms with dangling bonds, it would give opportunity to study influence 
of different atoms addition on them. For example addition of hydrogen atoms in case of graphene resulted in new material 
which is graphane ~\cite{grapha1}. \newline
Graphene and other nano-scale materials are recognized as future building blocks of new electronics technologies
~\cite{nano1}, including spintronics ~\cite{spin1}. In the case of low (one- and two-) dimensional structures  problem arises
because of famous Mermin-Wagner theorem ~\cite{mermin1}, which prevents ferro- or antiferromagnetic order to occur in finite 
temperatures, which is essential for practical application. This started the theoretical  and experimental search for magnetism 
in graphene and other two-dimensional structures.  One of the most promising directions is emergence of magnetism in such structures 
as an effect of presence of local defects ~\cite{exp1}. According to works of Palacios et al. ~\cite{palacios1} and, independently, 
of Yazyev ~\cite{yazyev1} single-atom defects can induce ferromagnetism in graphene based materials. 
In both cases, the magnetic order arises as an effect of presence of single-atom defects in combination with a sublattice
discriminating mechanism. In the case of \lin role of such defect could play non-hydrogenated N atom in hydrogenated structure. 
It would be then instructive to check influence of hydrogenation level on magnetic moment of the structure. \newline
In this paper electronic and magnetic structure of pure and hydrogenated 2DLi$_{2}$N have been analyzed by means of $ab$-$initio$ calculations.

\section{COMPUTATIONAL DETAILS}
To investigate electronic and magnetic properties of two-dimensional Li$_{3}$N structures a series of $ab$-$initio$ calculations 
have been conducted with use of DFT VASP code ~\cite{vasp1,vasp2} with PAW potentials ~\cite{vasp3}. For both spin-unpolarized and 
spin-polarized cases exchange-correlation potential has been approximated by generalized gradient approximation (GGA) using PW91 
functional ~\cite{pw91}. Kinetic energy cutoff of 500 eV for plane-wave basis set has been used. In all cases for self-consistent 
structure optimizations, the Brillouin zone (BZ) was sampled by $20\times 20\times 1$ special k points. All structures have been optimized for both, 
spin-unpolarized and spin-polarized cases unless Feynman-Hellman forces acting on each atom become smaller than 10$^{-4}$ eV/$\angstr$. 
A vacuum spacing of 12 \angstr \ was applied to hinder the interactions between \lin monolayers in adjacent cells. (dop. kiedy supercell i jak liczone magn.)  
Bandstructure and density of states (DoS) calculations have been confirmed by use of WIEN2k code ~\cite{wien1} which implements the full-potential 
linearized augmented plane wave (FLAPW) method \cite{flapw1}. In this case for exchange and correlation generalized gradient approximation 
was used in the Perdew-Burke-Ernzerhoff (PBE) parameterization \cite{gga1}. 

\section{RESULTS - ELECTRONIC STRUCTURE}
\begin{figure}[H]
  \begin{center}
    \begin{tabular}{cc}
      \resizebox{90mm}{!}{\includegraphics[angle=0]{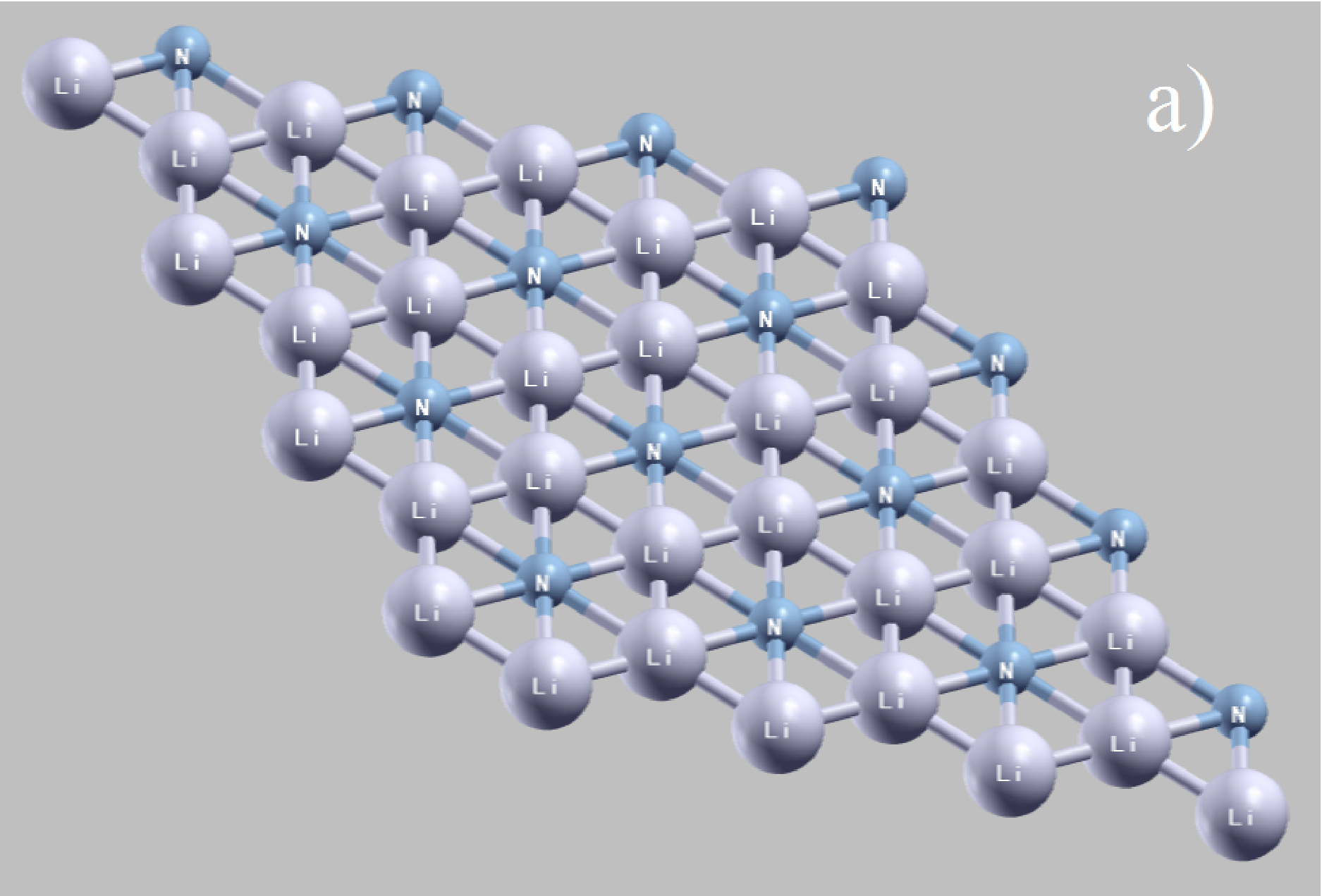}} &
			\resizebox{90mm}{!}{\includegraphics[angle=0]{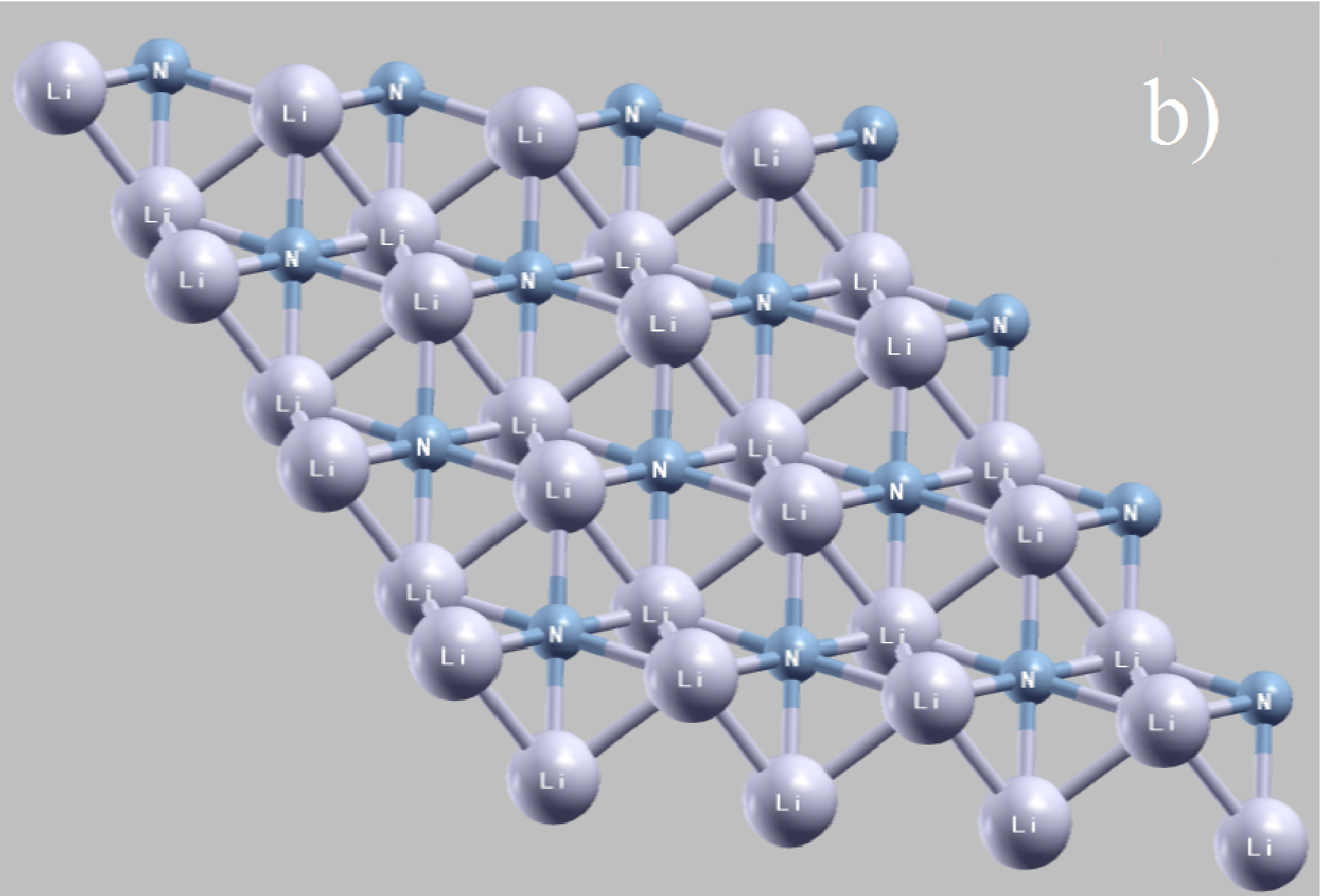}} 
    \end{tabular}
    \caption{Atomic structure of plane (a) vs. low-buckled (b) \lin \ (color online)}
    \label{fig0}
  \end{center}
\end{figure}

\begin{figure}[H]
  \begin{center}
    \begin{tabular}{cc}
      \resizebox{90mm}{!}{\includegraphics[angle=0]{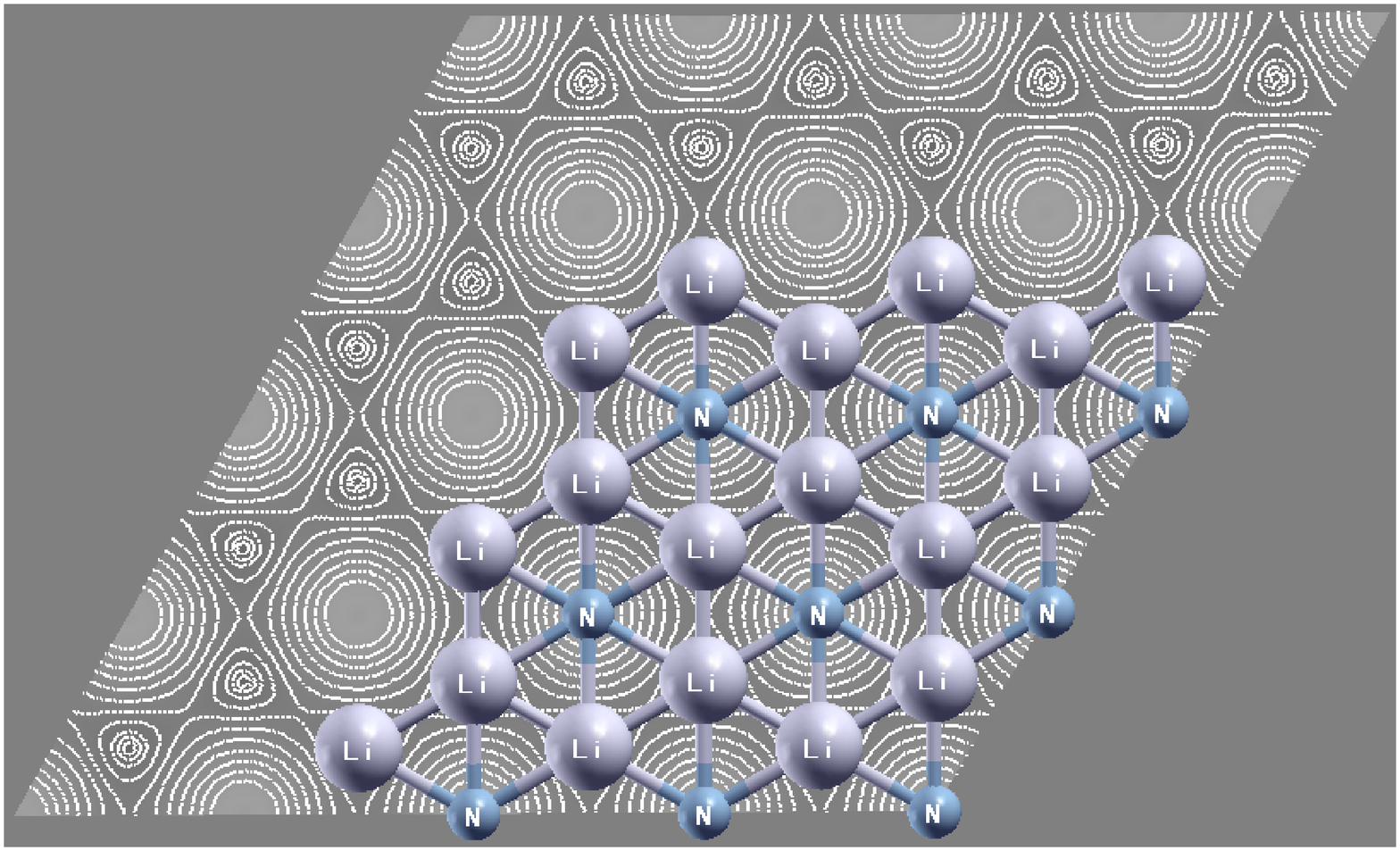}} &
    \end{tabular}
    \caption{Valence charge density of (?) (color online)}
    \label{fig1}
  \end{center}
\end{figure}

To study electronic properties of \lin, at first comparison has been made with 
bulk material. For both cases lattice constants have been determined
by total energy calculations and are found to be equal to 3.65 \angstr \ for 
bulk (experimental value 3.63 \angstr) and 3.57 \angstr \ for \lin. 
In agreement with~\cite{lin4} bulk Li$_{3}$N is a semiconductor with non-direct 
bandgap equal to 1.15 eV between A (valence band) and $\Gamma$ (conduction band)
points. In contradiction to this, \lin \ has metallic nature. \newline 
Two-dimensional structure is rather weakly bound - binding energy 
(defined as $E_{b} = E_{at} - E_{sheet}$ where $E_{at}$ is
the energy of isolated atom(s) and $E_{sheet}$ is the total energy of 
two-dimensional structure) is equal to 10.36 eV, while binding energy of bulk 
structure is equal to 14.25 eV. Also, two dimensional sheet would have N atoms 
with dangling bonds, such structure would be then rather unstable 
with respect to foreign atoms addition. Graphane case suggests that it 
would be instructive to examine influenece of hydrogenation 
on electronic structure in such cases as well as addition of lithium atoms. \newline
The nature of Li-N bond is ionic, as it can be seen from Fig. \ref{fig1} showing charge density
projected on [110] plane. Since every bond has both ionic and covalent character
the level of ionicity can be estimated using difference between electronegativities of 
bonded atoms ~\cite{pauling}. In the case of Li-N bond this difference equal to 2 suggests, that 
the bond is about 65$\%$ ionic and 35$\%$covalent. This fact together with rather large lattice 
constant suggest that the structure of two-dimensional \lin \ can be low-buckled (LB) rather than plane (PL),   
according to puckering mechanism described in ~\cite{III-V1}. To check this the series of calculations
has been done, each with different distance in z direction between Li atoms and the plane on which N atoms lie ($\Delta z$).
The structure with minimal energy has been then optimized. Calculations show, that the buckled structure with $\Delta z$ = X lies 0.54 eV 
lower that the plane, which means that indeed the puckering mechanism stabilizes the structure.
Both, plane and low-buckled structures can be seen on Fig \ref{fig0}. \newline
Four structures have been then studied in two conformations, plane and low-buckled --- two (PL and LB) with single H atom 
attached on top of each N atom (\linh), two with two H atoms attached on both sides of N (\linhh), 
two with single Li atom attached on top of each N atom (\linli), and two with two Li atoms attached on both 
sides of N (\linlili). 

Calculated binding energies, hydrogen-addition energies, lithium-addition energies, and 
structure parameters are shown in Tab. 1 (plane structure) and  Tab. 2 (low-buckled structure). In Tab. 1  
d$_{N-H}$ is the distance between N and  passivated H atoms, d$_{N-Li}$ is the distance between N and passivated Li atoms, and 
$\Delta z$ is distance in the $z$ direction between Li atoms and the plane on which N atoms lie. In tab. 2 E$_{t}$ is total energy 
of the structure with reference to plane structure. As can be seen, in all cases low-buckled structure is lower in energy, from 0.01 eV
for \linlili \ to 1.08 eV for \linhh. So, from now on, all properties are referred to LB structure, unless stated otherwise.

\begin{table}[H]
\begin{tabular}{lccccccccc}
\hline
\hline
                             & 2DLi$_{2}$N & \ & 2DLi$_{2}$N+H & \ & 2DLi$_{2}$N+2H & \ & 2DLi$_{2}$N+Li & \ & 2DLi$_{2}$N+2Li \\  \hline
E$_{b}$ (eV)                 & 10.36       & \ & 15.98         & \ & 16.88          & \ & 13.04          & \ & 15.37           \\ 
E$_{add}$ (eV)               & -           & \ & 5.80          & \ & 0.90           & \ & 2.86           & \ & 2.29            \\
d$_{N-H}$ (\angstr)          & -           & \ & 1.046         & \ & 1.157          & \ & -              & \ & -               \\ 
d$_{N-Li}$ (\angstr)         & -           & \ & -             & \ & -              & \ & 1.88           & \ & 1.99            \\ 
$\Delta$z$_{N-Li}$ (\angstr) & 0.0         & \ & 0.129         & \ & 0.0            & \ & 0.187          & \ & 0.0             \\  \hline
\end{tabular}
\caption{Comparison of binding and addition energies and structure parameters of two-dimensional plane \lin. }  
\end{table}

\begin{table}[H]
\begin{tabular}{lccccccccc}
\hline
\hline
                             & 2DLi$_{2}$N & \ & 2DLi$_{2}$N+H & \ & 2DLi$_{2}$N+2H & \ & 2DLi$_{2}$N+Li & \ & 2DLi$_{2}$N+2Li \\  \hline
E$_{t}$ (eV)                 & 0.54        & \ & 0.29          & \ & 1.08           & \ & 0.21           & \ & 0.01            \\ 
E$_{add}$ (eV)               & -           & \ & 5.55          & \ & 1.44           & \ & 2.11           & \ & 1.74            \\
d$_{N-H}$ (\angstr)          & -           & \ & 1.05          & \ & 1.21           & \ & -              & \ & -               \\ 
d$_{N-Li}$ (\angstr)         & -           & \ & -             & \ & -              & \ & 1.91           & \ & 2.00            \\ 
$\Delta$z$_{N-Li}$ (\angstr) & 0.49        & \ & 0.60          & \ & 0.84           & \ & 0.30           & \ & 0.22            \\  \hline
\end{tabular}
\caption{Comparison of total energy (with reference to plane structure) and structure parameters of two-dimensional low-buckled \lin. }  
\end{table}

Binding energy of \linh, equal to 16.27 eV, is comparable to binding energy of 
bulk structure (both have the same number of atoms in the unit cell), which suggests
that H-passivated two-dimensional structure would be no less stable than the bulk.
Addition energies are defined as 
$E_{add}=E_{tot}(\mbox{layer+addatom})-
 E_{tot}(\mbox{layer})-E_{tot}(\mbox{addatom})$.
High value of addition energy for single H atom equal to 5.55 eV suggests, that
hydrogen addition may stabilize the structure, making it more bound.
Addition of second H atom to already H-passivated structure requires only 1.44 eV,
which means that \linhh \ structure is less stable than \linh. 
Addition energy of 2.11 eV for single Li atom on pure layer is almost equal to addition 
energy for single Li atom on Li-passivated structure (1.74 eV), but such structures 
are less bound that H-passivated cases. \newline

MW $\rightarrow$ \newline
The electronic structure of a single-layer \lin\ (not taking into account the Fermi level) 
exhibits typical structure of a semiconductor, with valence and conduction bands
separated by a band gap of 4 eV. Its metallic character is purely due to the position of
Fermi level which is lowered by -2 eV from the middle of the band gap to the 
upper part of the valence band.
This may be explained by a large contribution to the density of states of the valence band
coming from the p-electrons from N atoms. 
The conduction band of a single-layer \lin\ originates
mainly from the density of states of p-electrons of lithium.

The relative positions of bands in band structure of \linh\ with single H atoms added,
remain unchanged with respect to \linh \, with the exception of 
a single nitrogen p-electrons band, which is lowered by 2 eV.
The Fermi level is rised by 2 eV from the conduction band to
the middle of the band gap, and therefore \linh\ is semiconductor. The rising of the Fermi level
is caused by an increased contribution to the density of states from lithium p-electrons 
\textcolor{red}{(see Fig. 2e)}. The contributions of s-states originating from additional hydrogen 
is insignificant.

The same effect can be observed when the second H atom is added to the \linh\ layer.
The Fermi level is rised again by 2 eV with respect to the position of bands, and
is placed at the bottom of the conduction band. Therefore \linhh\ is again metallic.

We can summarize the effect of H atom addition:
each H atoms rises the Fermi level by 2 eV and the band structure remains unchanged.

Addition of lithium atoms to the single-layer \lin\ leads to quite different effects.
The band structure of \linli\ is purely metallic with no band gap. 
The energy band of p-electrons for additional litium atom connects the conduction
band of the original \lin\ band structure at K and M points of the Brillouine
zone with the valence band at $\Gamma$ point, effectively nullyfing the energy gap.
There exists non-zero density of
states of p-electrons for additional litium atom directly at the Fermi level.
The Fermi level is practicaly unchanged with respect to the band structure and remains
at the top of the conduction band of the original \lin\ band structure.

Addition of second lithium atom repeats the effects of the first one. There appears 
a new steep band of lithium p-electrons and the resulting structure remains metallic.
The Fermi level is shifted up by about 1 eV with respect to the band structure, however
is does not influence the metallic character of the compound.

We can summarize the effect of Li atom additions: each Li atoms adds an energy band
directly across the energy gap of the the original \lin\ band structure. \newline

MW $\leftarrow$ \newline


\begin{figure}[H]
  \begin{center}
    \begin{tabular}{ccc}
      \resizebox{60mm}{!}{\includegraphics[angle=270, trim=0mm 45mm 0mm 55mm, clip=true]{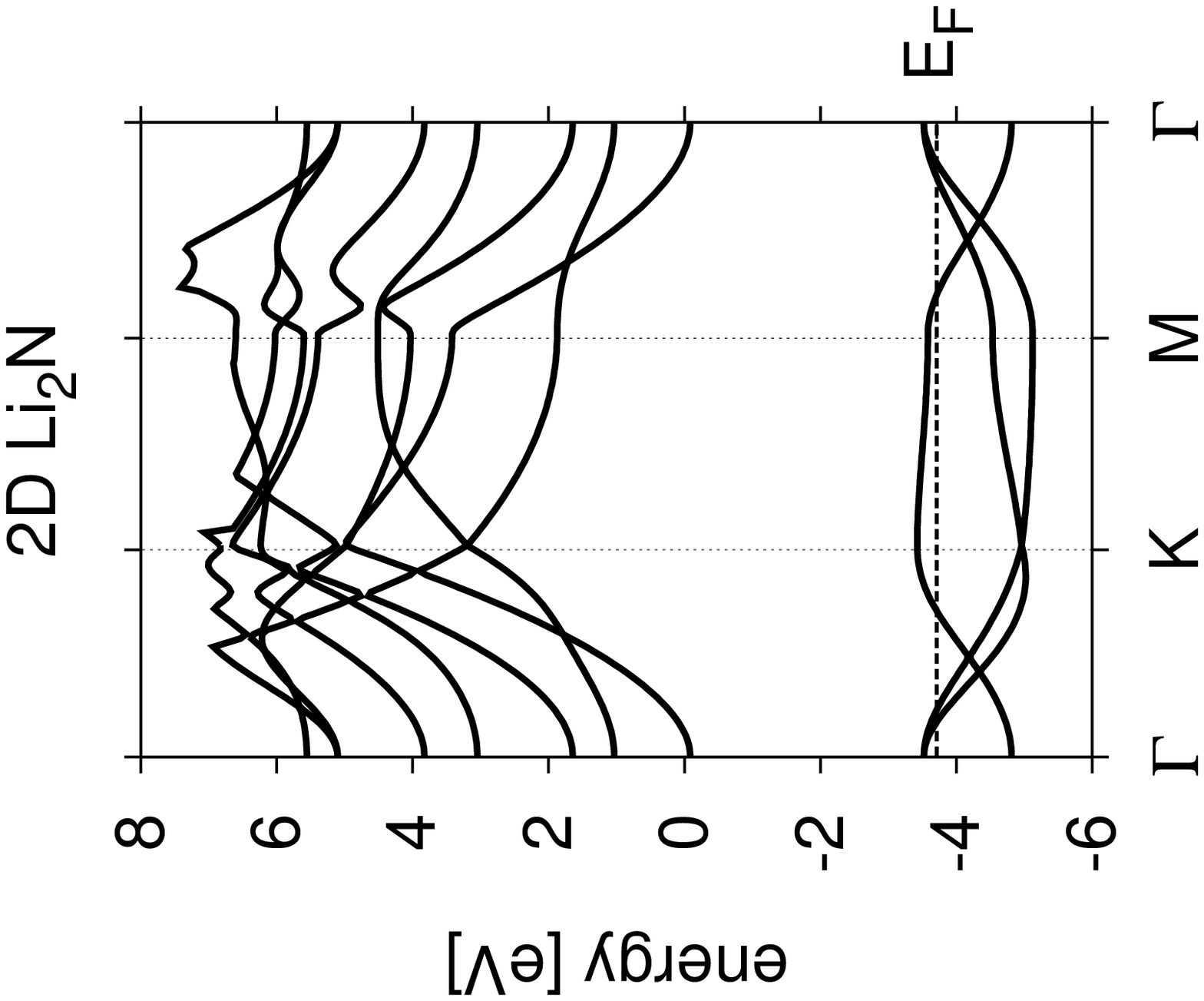}} &
			\resizebox{60mm}{!}{\includegraphics[angle=270, trim=0mm 45mm 0mm 55mm, clip=true]{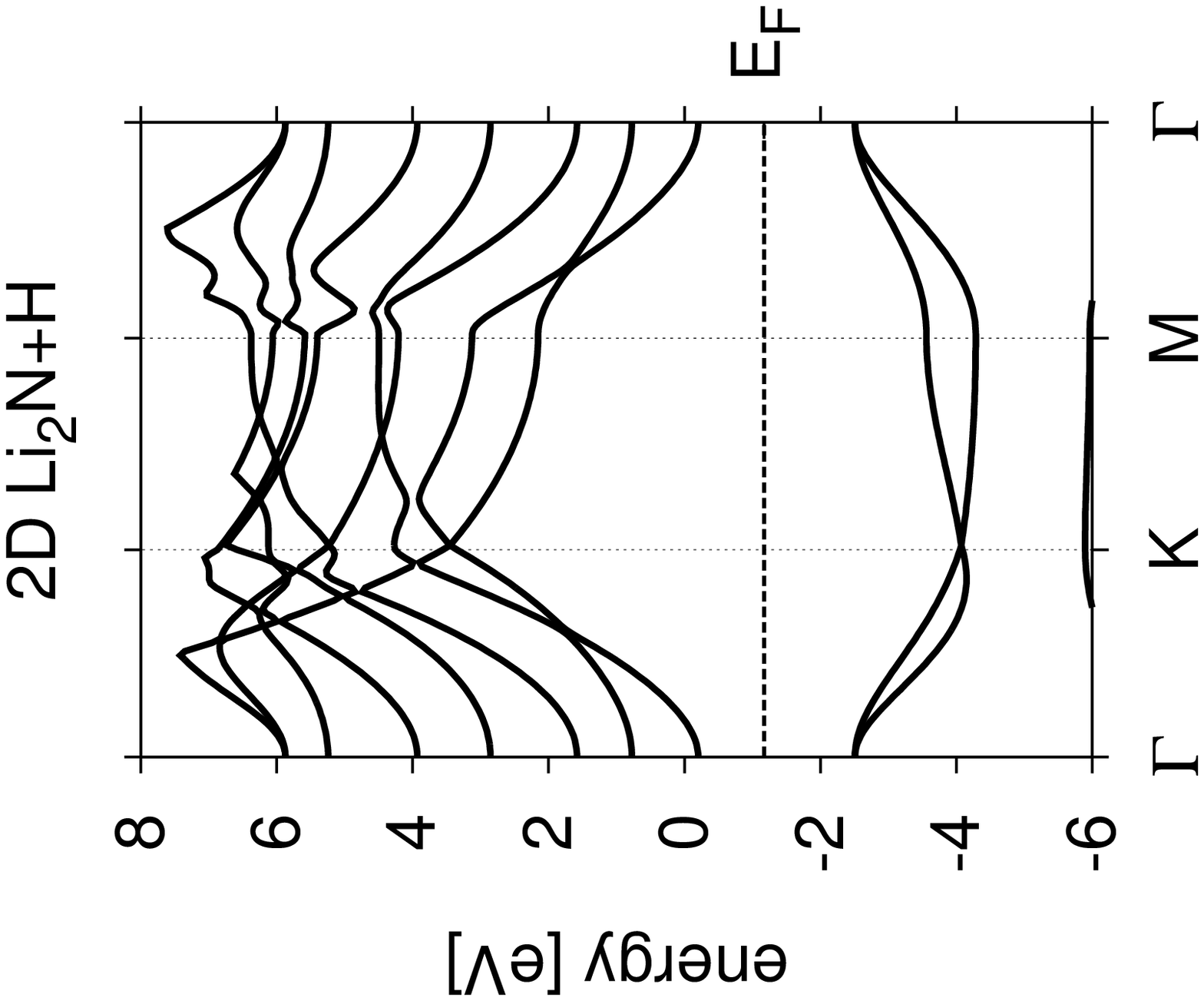}} &
      \resizebox{60mm}{!}{\includegraphics[angle=270, trim=0mm 45mm 0mm 55mm, clip=true]{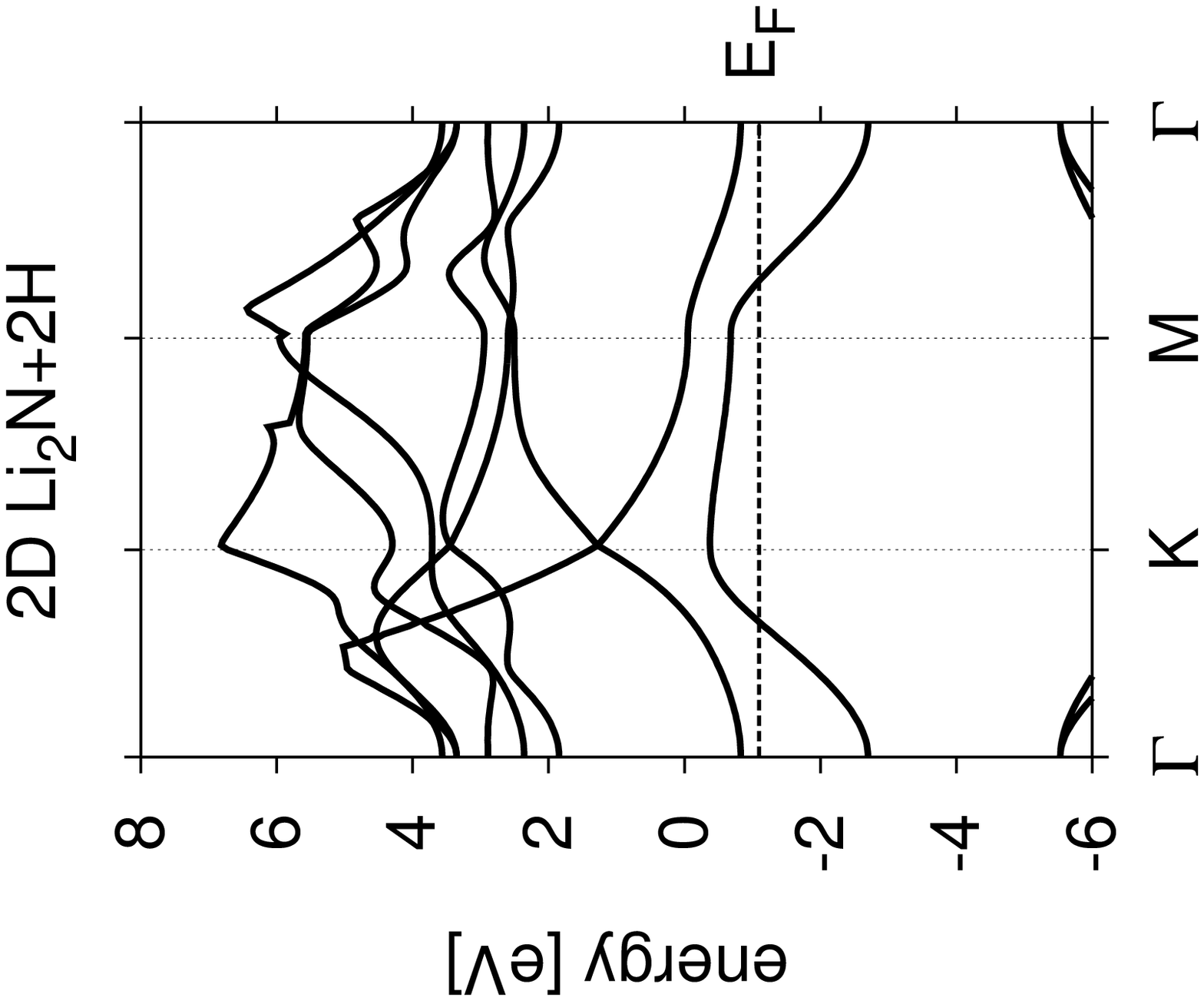}} \\
      \resizebox{60mm}{!}{\includegraphics[angle=270, trim=0mm 45mm 0mm 55mm, clip=true]{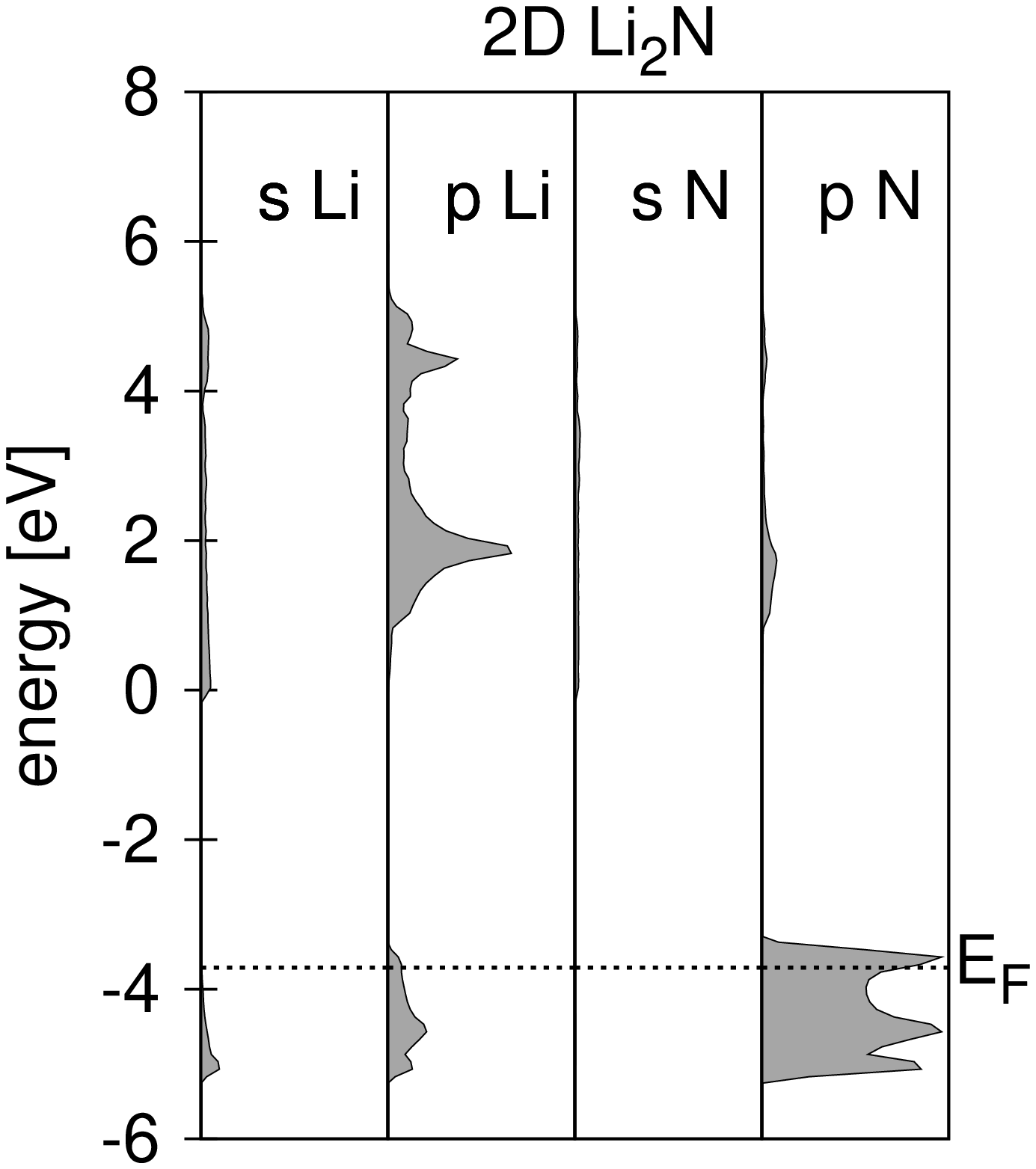}} &
			\resizebox{60mm}{!}{\includegraphics[angle=270, trim=0mm 45mm 0mm 55mm, clip=true]{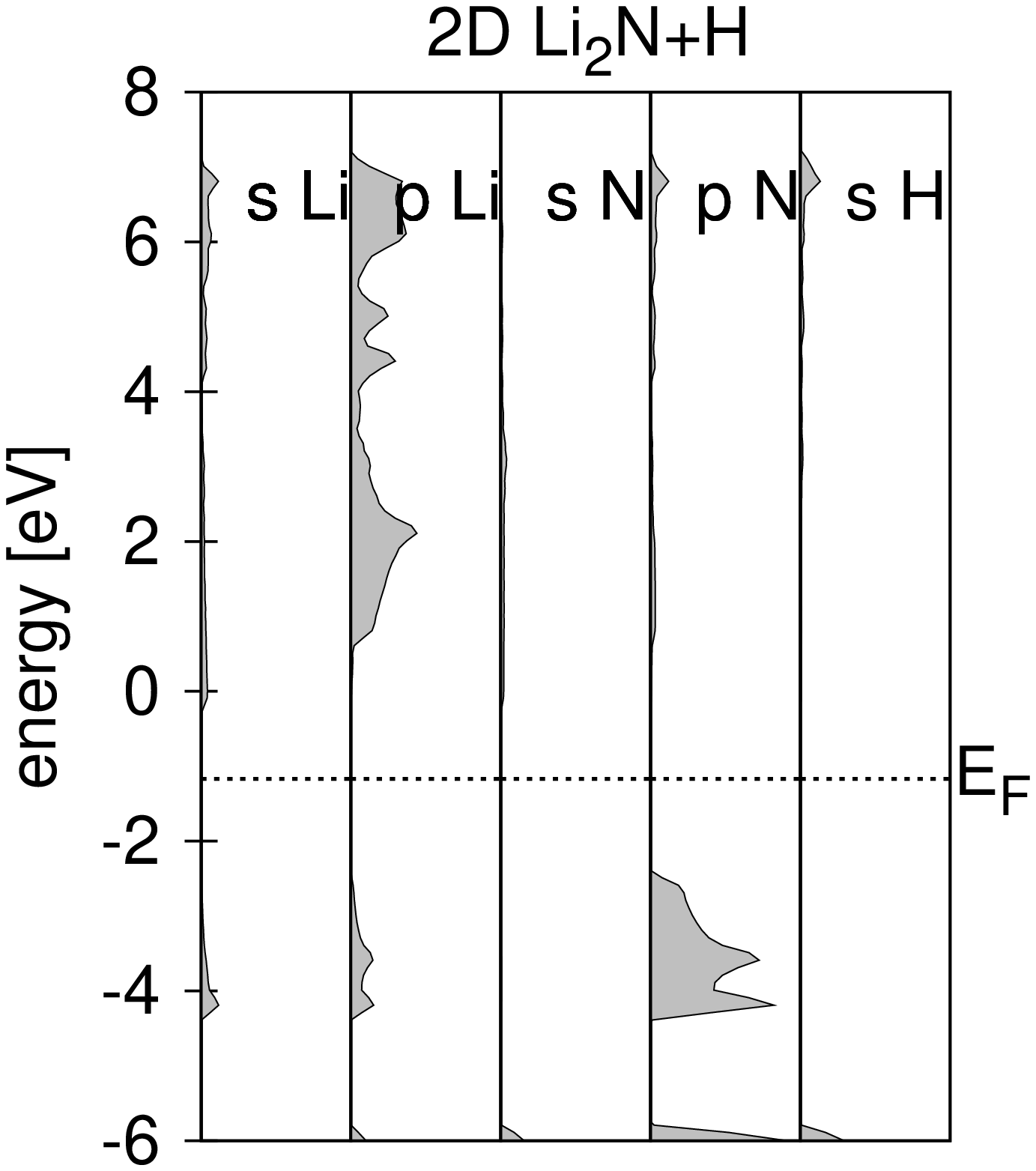}} &
      \resizebox{60mm}{!}{\includegraphics[angle=270, trim=0mm 45mm 0mm 55mm, clip=true]{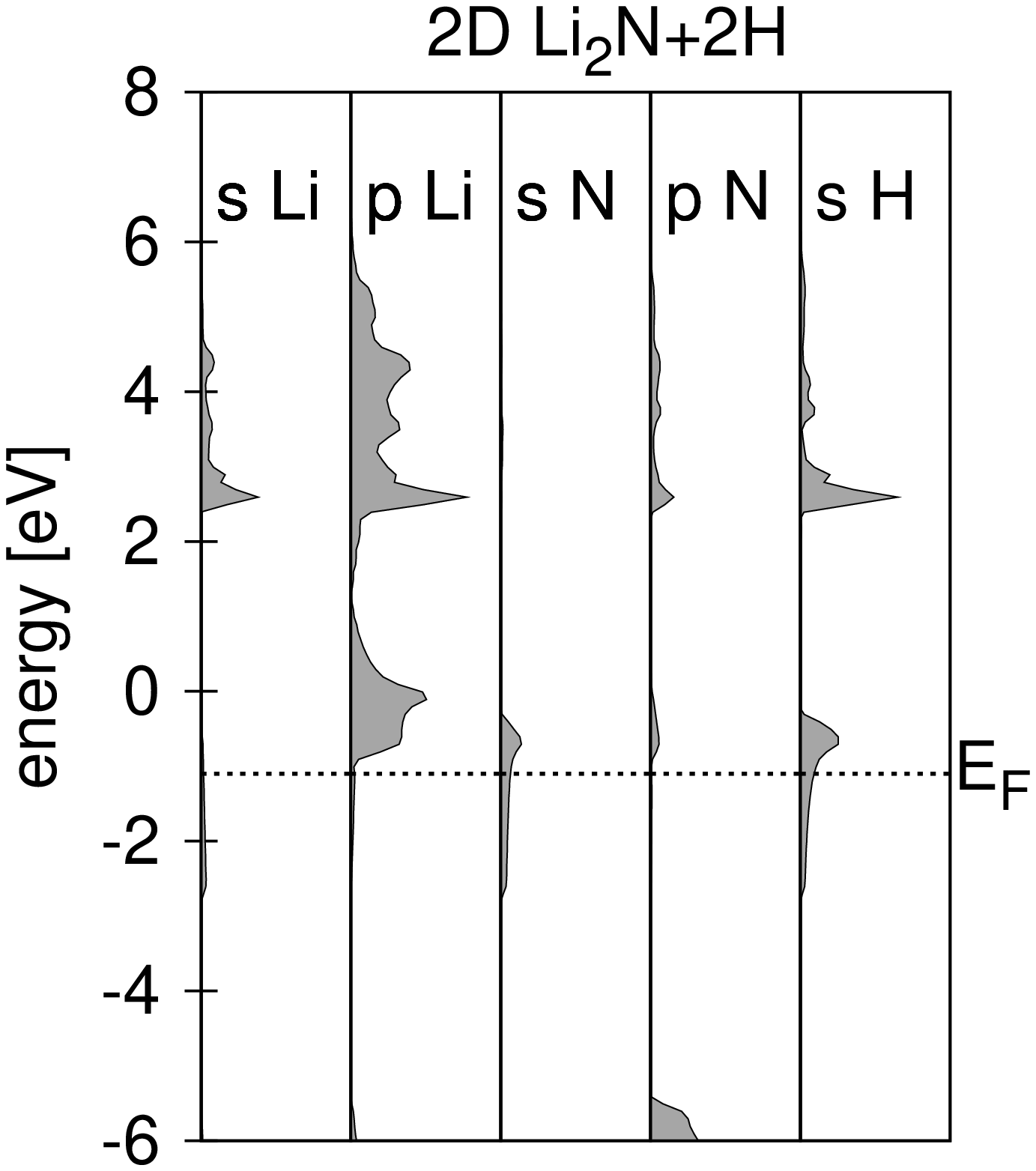}} \\
    \end{tabular}
    \caption{Bandstructures and partial density of states for PL \lin, \linh \ and \linhh. Details in text.}
    \label{fig2}
  \end{center}
\end{figure}

\begin{figure}[H]
  \begin{center}
    \begin{tabular}{ccc}
      \resizebox{60mm}{!}{\includegraphics[angle=270, trim=0mm 45mm 0mm 55mm, clip=true]{bands_Li2N.ps}} &
			\resizebox{60mm}{!}{\includegraphics[angle=270, trim=0mm 45mm 0mm 55mm, clip=true]{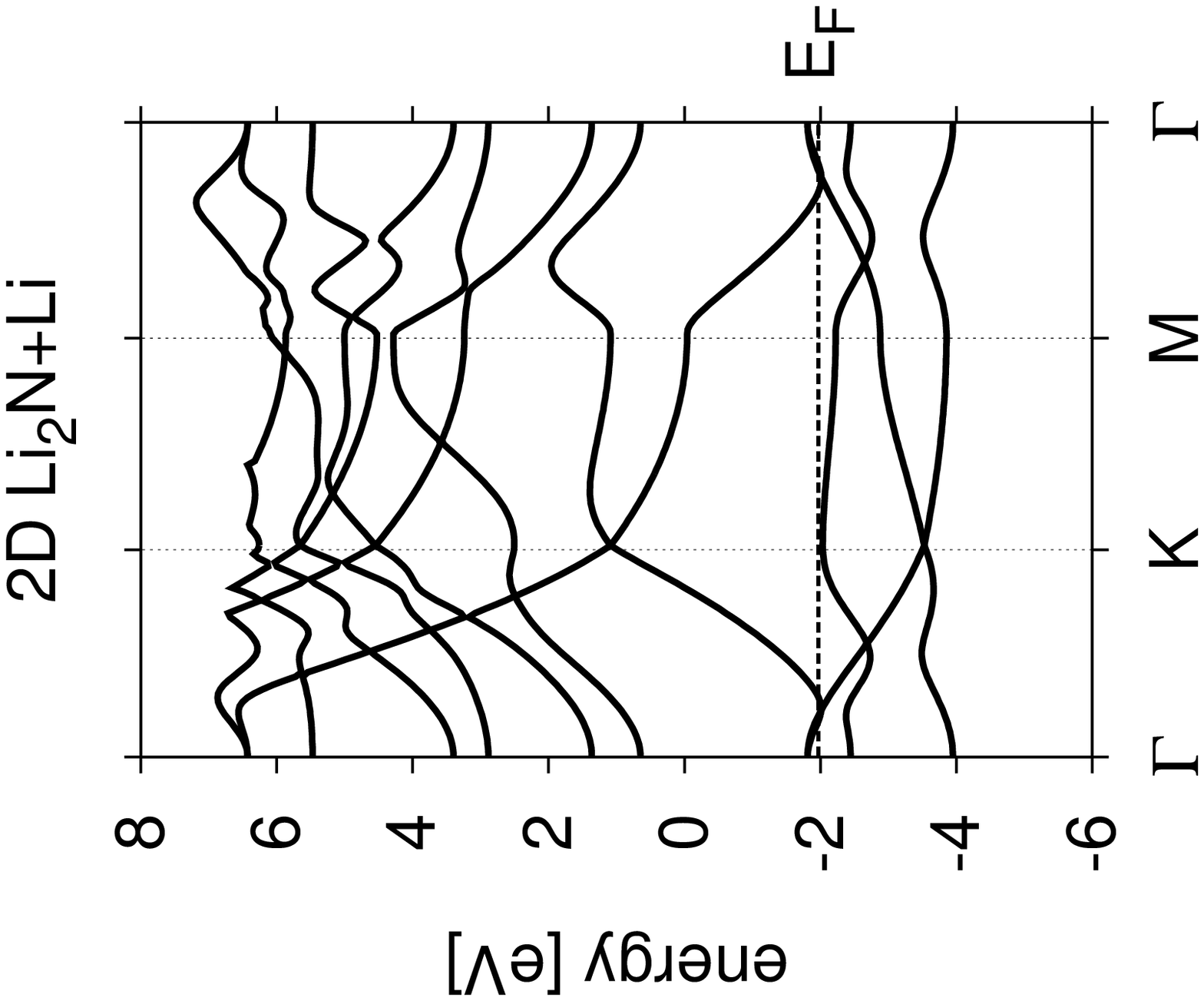}} &
      \resizebox{60mm}{!}{\includegraphics[angle=270, trim=0mm 45mm 0mm 55mm, clip=true]{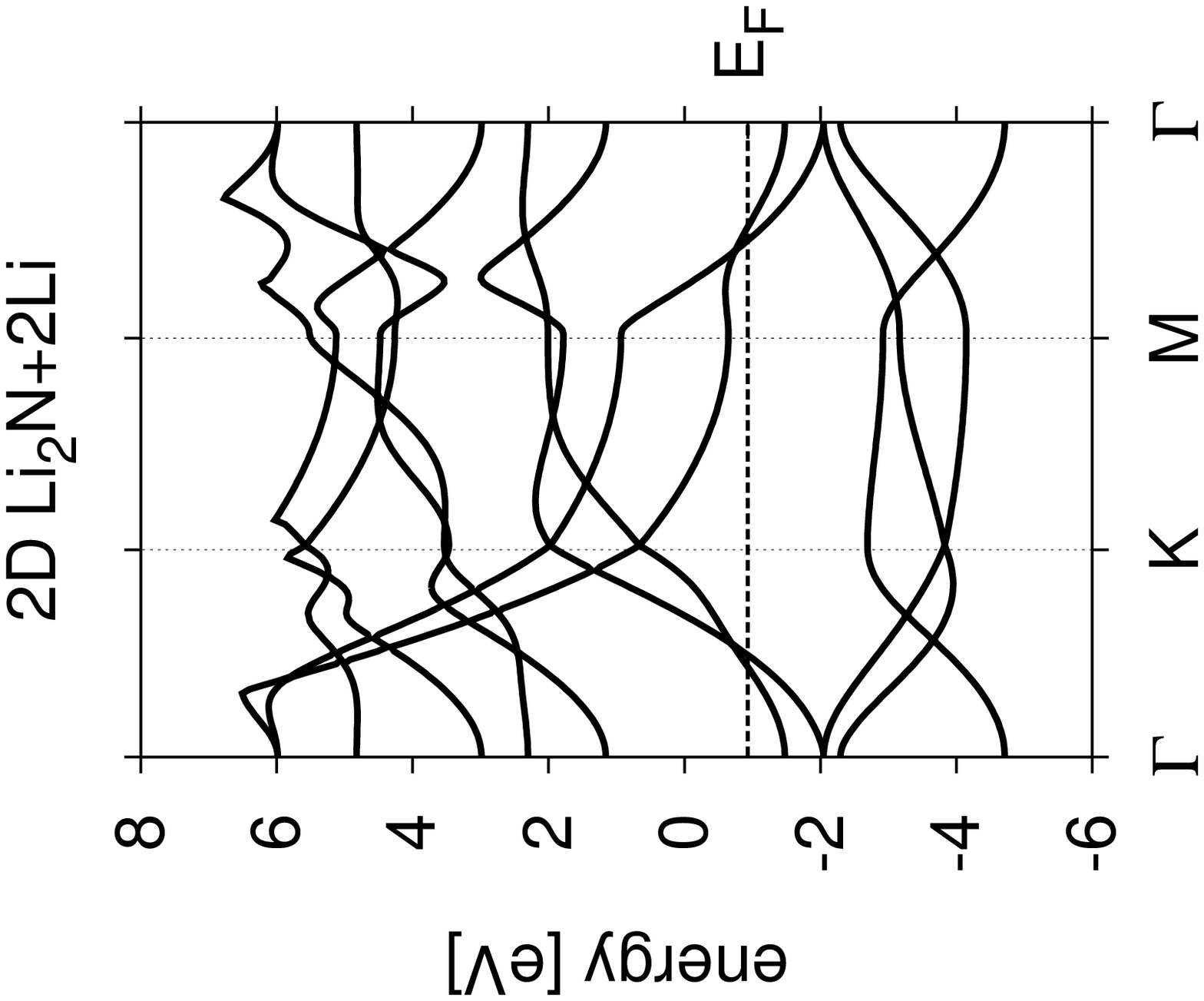}} \\
      \resizebox{60mm}{!}{\includegraphics[angle=270, trim=0mm 45mm 0mm 55mm, clip=true]{pdos_Li2N.ps}} &
			\resizebox{60mm}{!}{\includegraphics[angle=270, trim=0mm 45mm 0mm 55mm, clip=true]{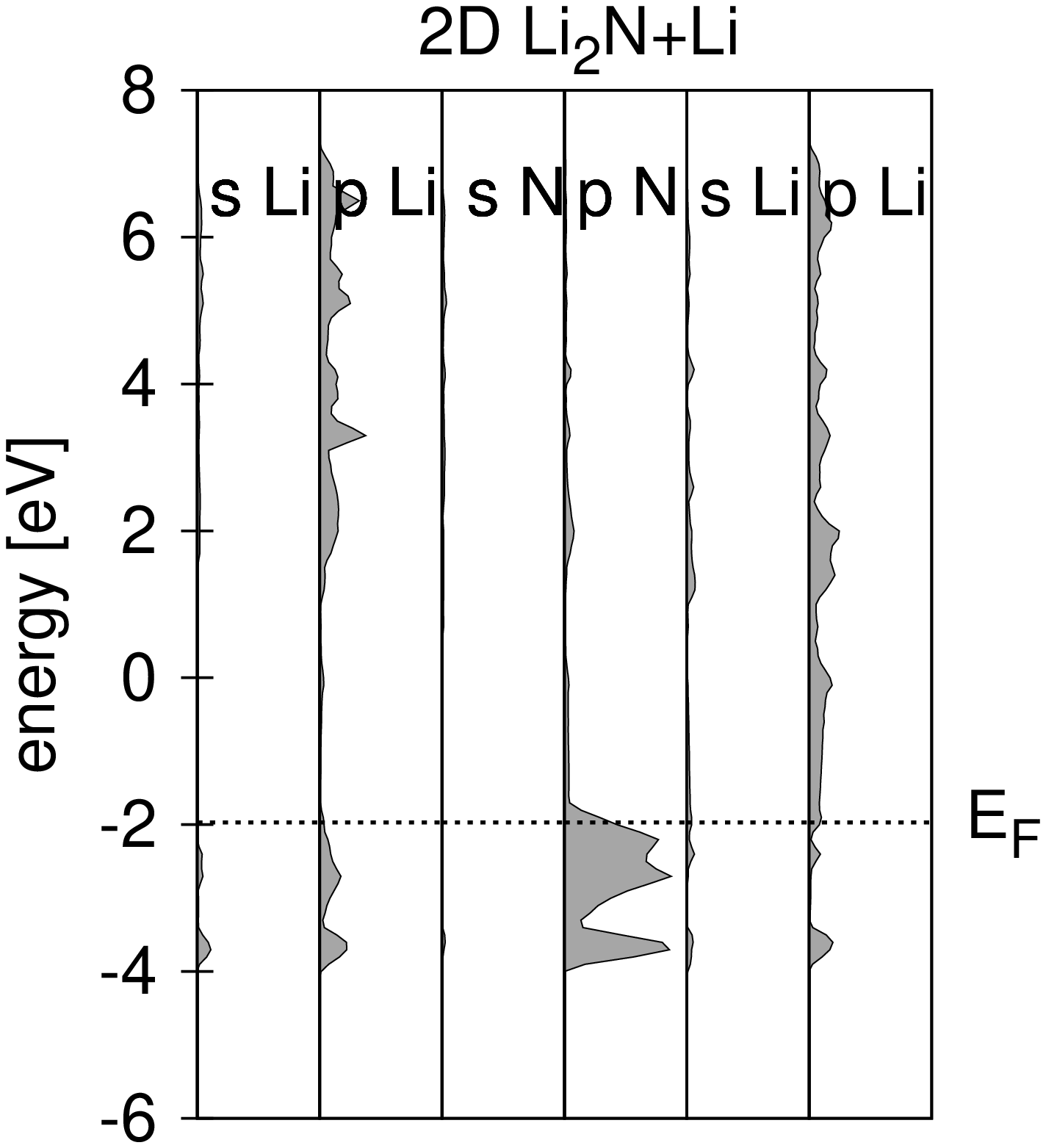}} &
      \resizebox{60mm}{!}{\includegraphics[angle=270, trim=0mm 45mm 0mm 55mm, clip=true]{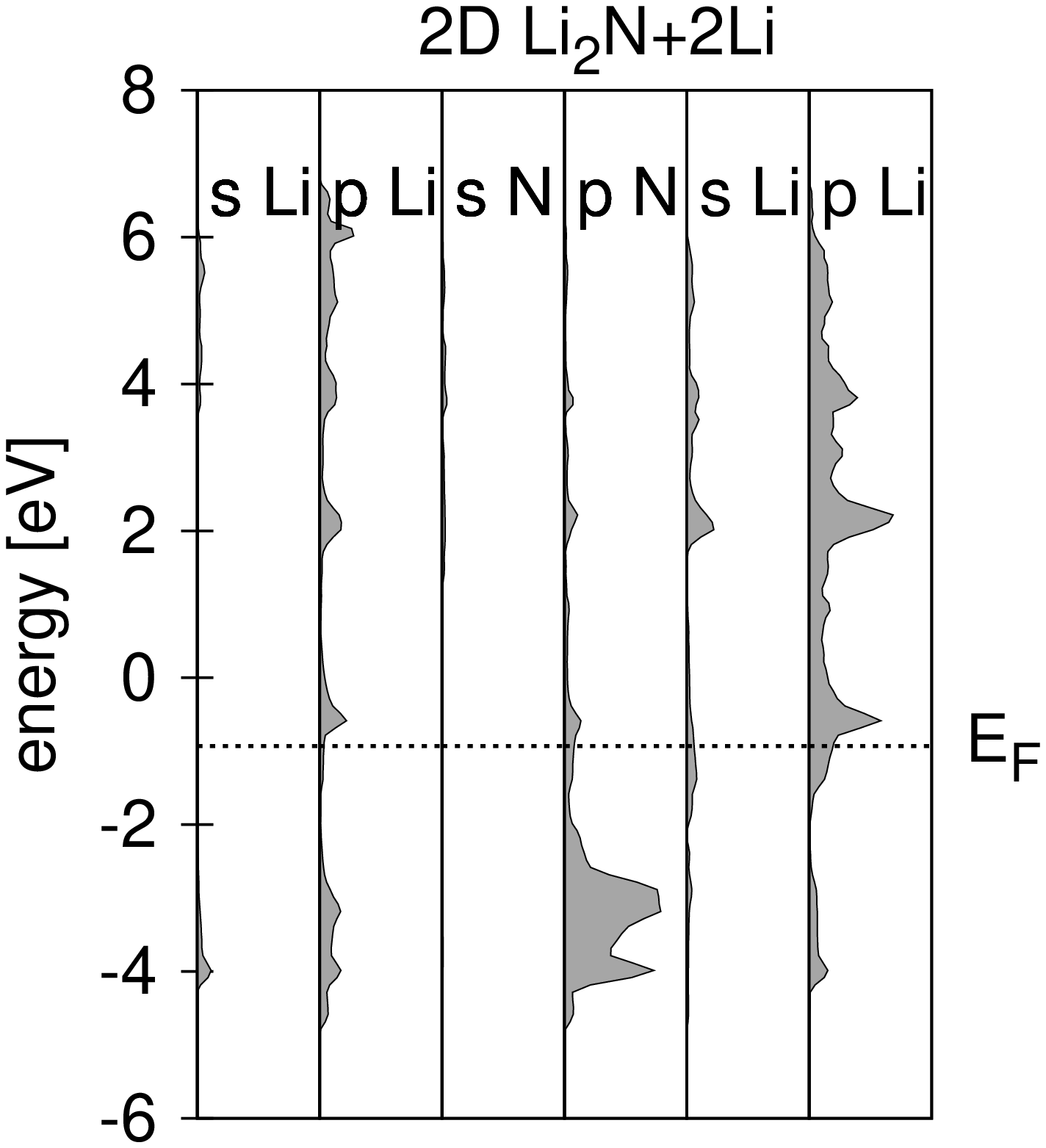}} \\
    \end{tabular}
    \caption{Bandstructures and partial density of states for PL \lin, \linli \ and \linlili. Details in text.}
    \label{fig3}
  \end{center}
\end{figure}

\begin{figure}[H]
  \begin{center}
    \begin{tabular}{ccc}
      \resizebox{60mm}{!}{\includegraphics[angle=270, trim=0mm 45mm 0mm 55mm, clip=true]{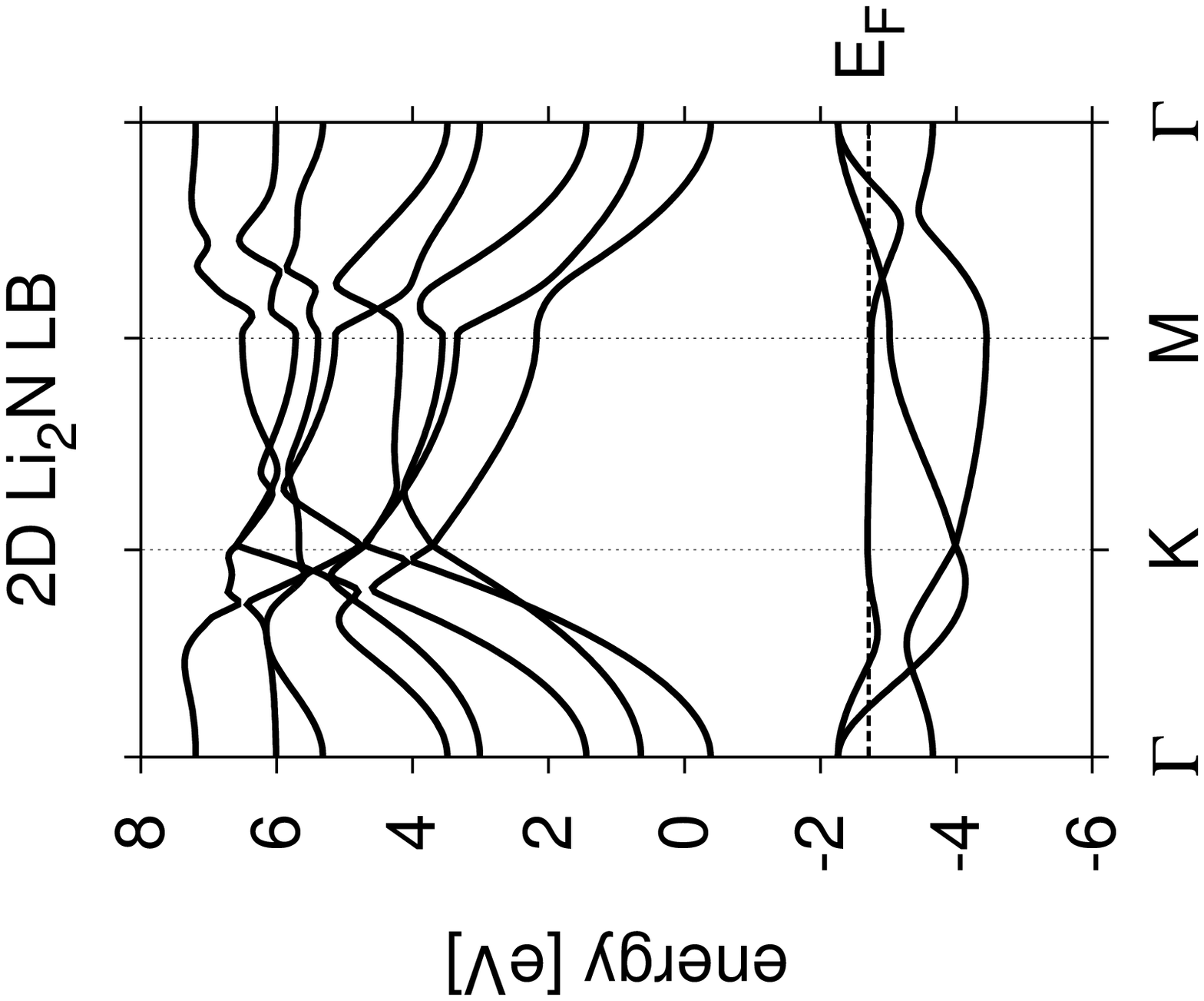}} &
			\resizebox{60mm}{!}{\includegraphics[angle=270, trim=0mm 45mm 0mm 55mm, clip=true]{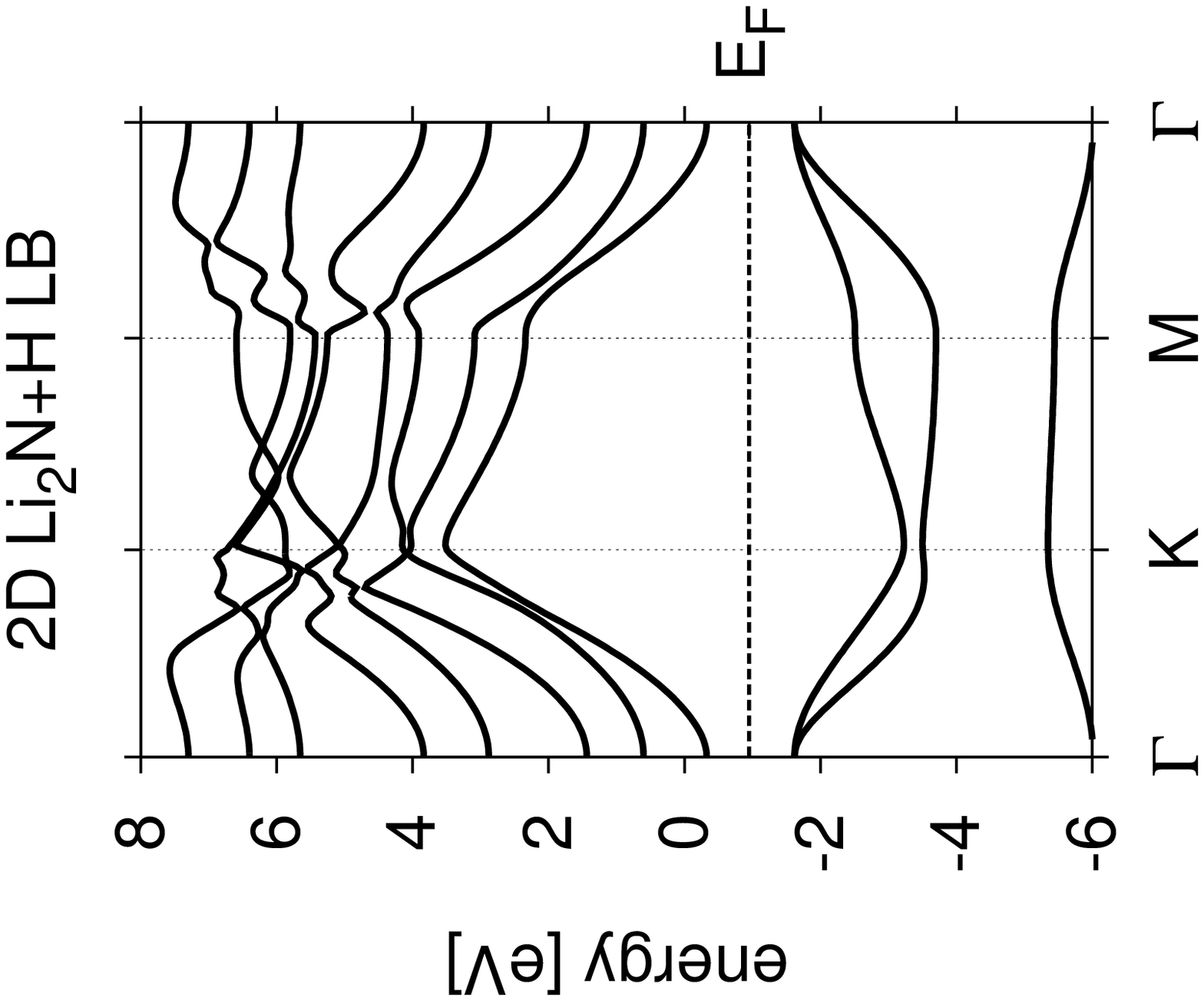}} &
      \resizebox{60mm}{!}{\includegraphics[angle=270, trim=0mm 45mm 0mm 55mm, clip=true]{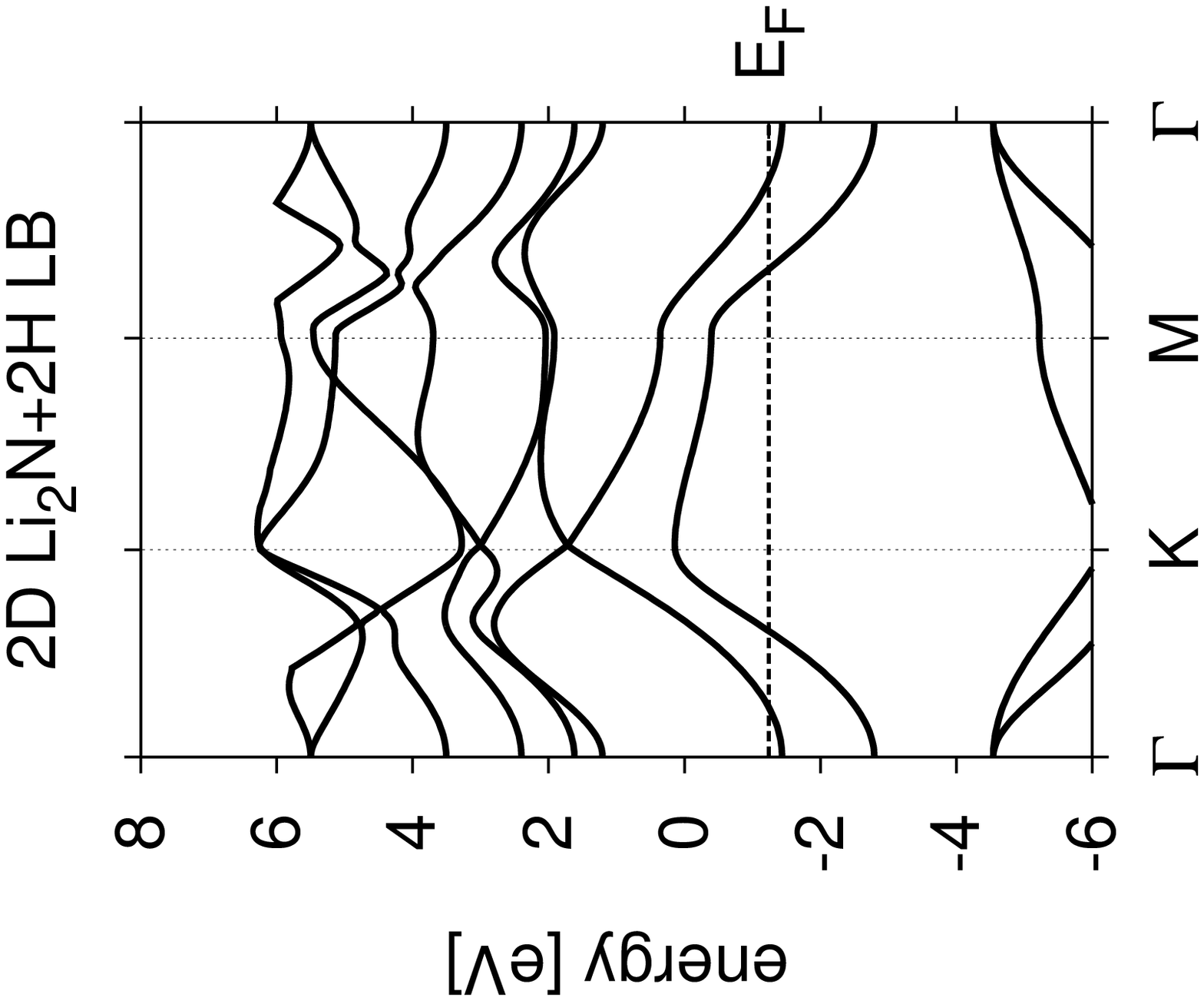}} \\
      \resizebox{60mm}{!}{\includegraphics[angle=270, trim=0mm 45mm 0mm 55mm, clip=true]{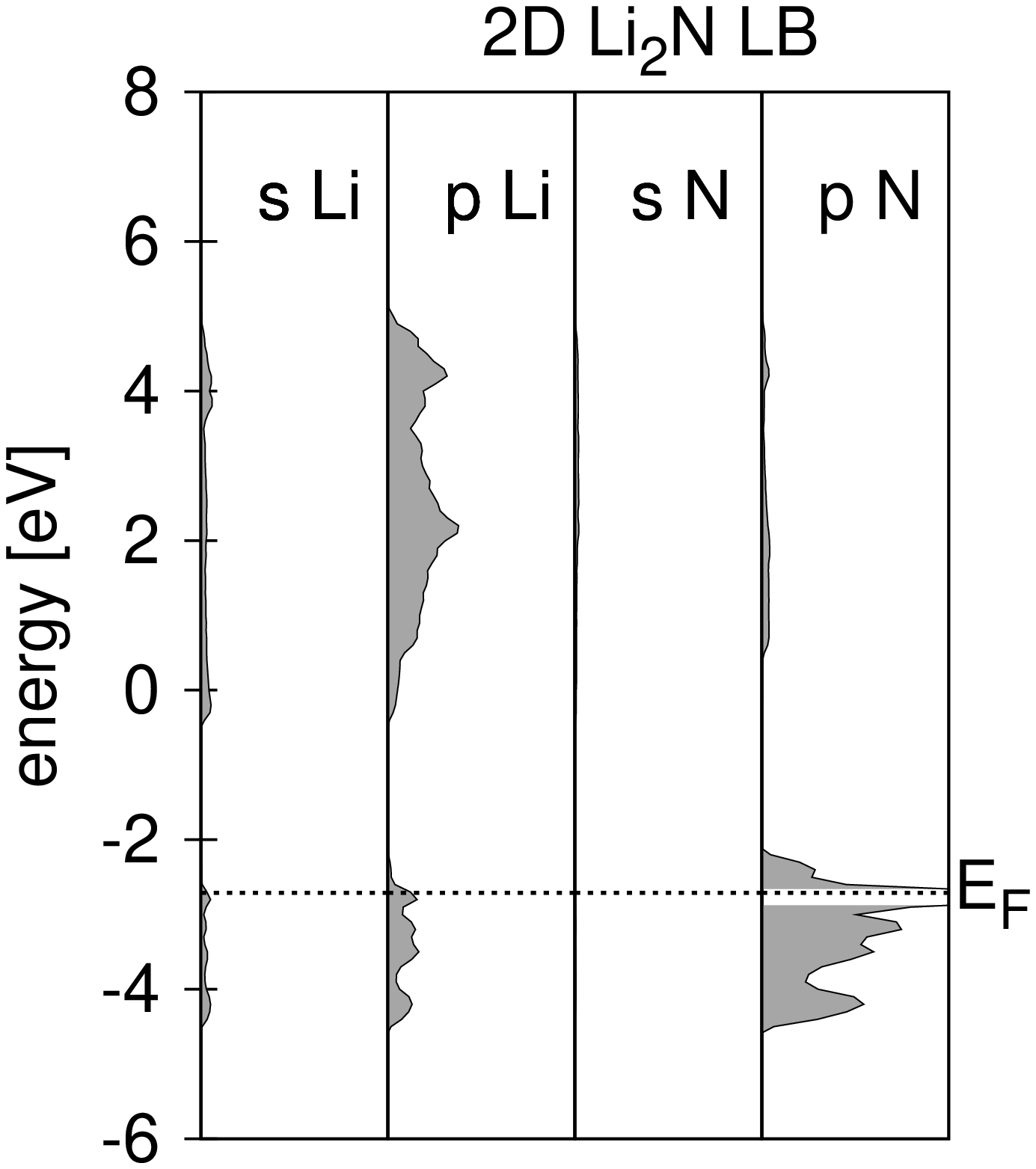}} &
			\resizebox{60mm}{!}{\includegraphics[angle=270, trim=0mm 45mm 0mm 55mm, clip=true]{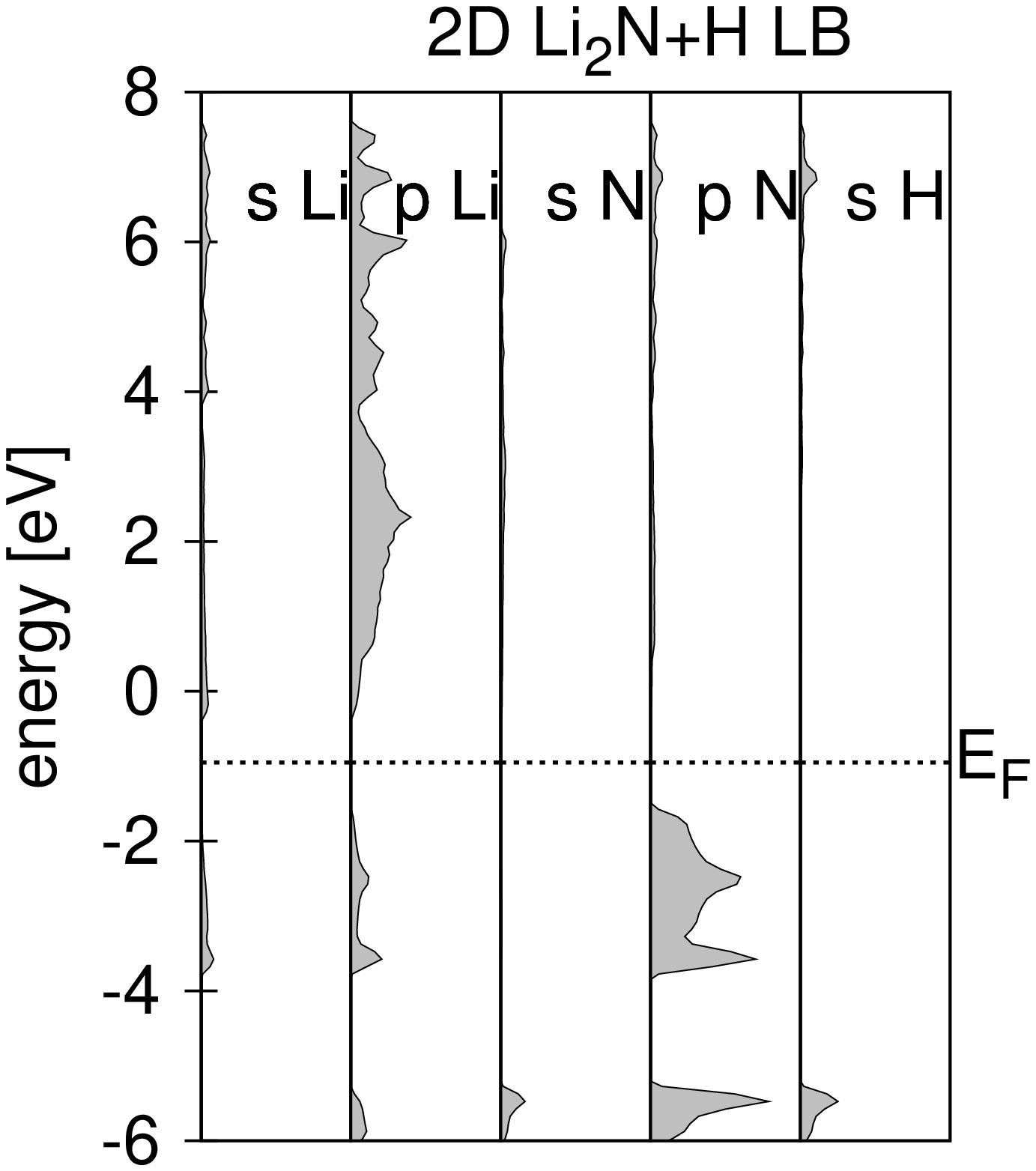}} &
      \resizebox{60mm}{!}{\includegraphics[angle=270, trim=0mm 45mm 0mm 55mm, clip=true]{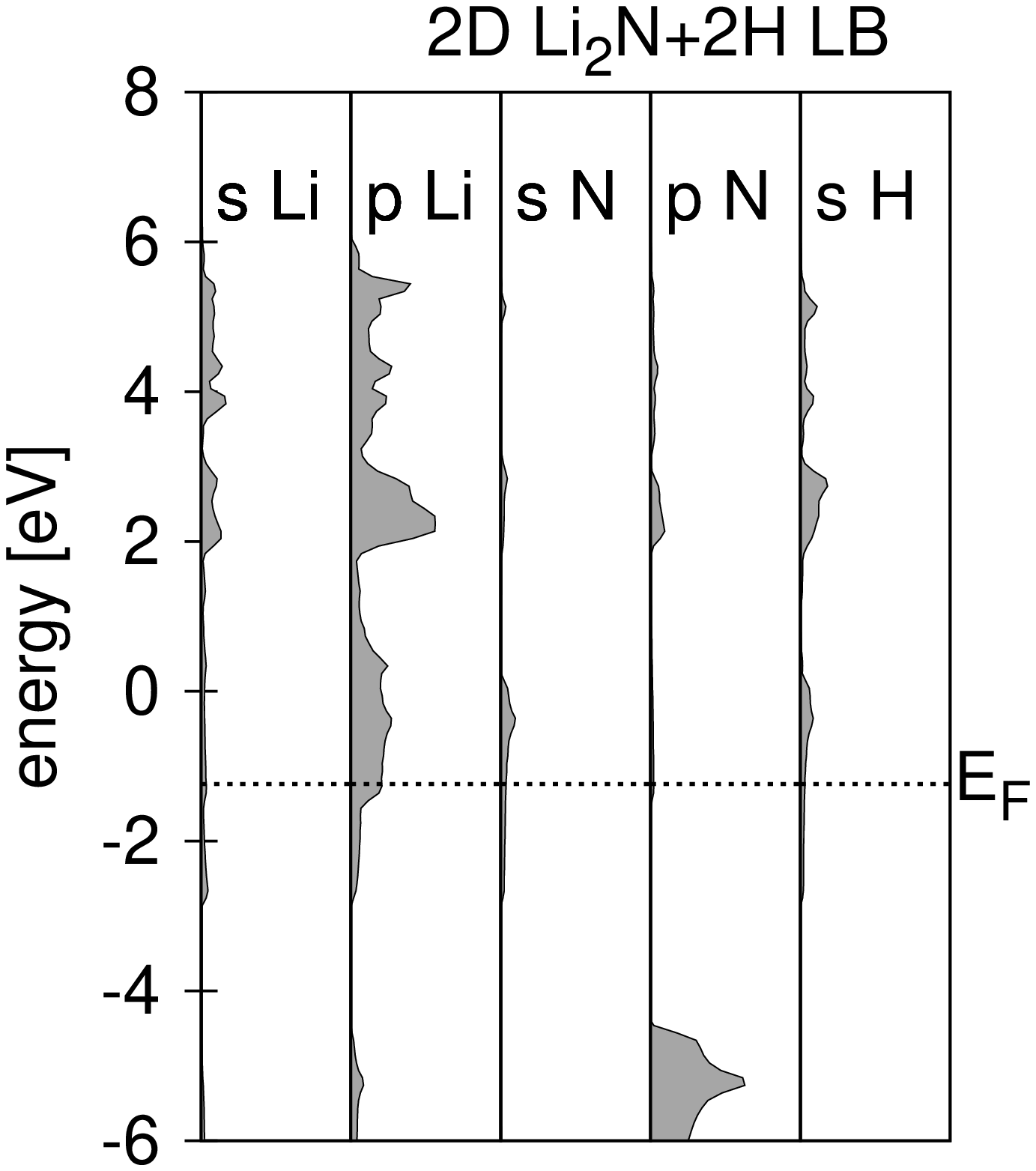}} \\
    \end{tabular}
    \caption{Bandstructures and partial density of states for LB \lin, \linh \ and \linhh. Details in text.}
    \label{fig4}
  \end{center}
\end{figure}

\begin{figure}[H]
  \begin{center}
    \begin{tabular}{ccc}
      \resizebox{60mm}{!}{\includegraphics[angle=270, trim=0mm 45mm 0mm 55mm, clip=true]{bands_Li2N_LB.ps}} &
			\resizebox{60mm}{!}{\includegraphics[angle=270, trim=0mm 45mm 0mm 55mm, clip=true]{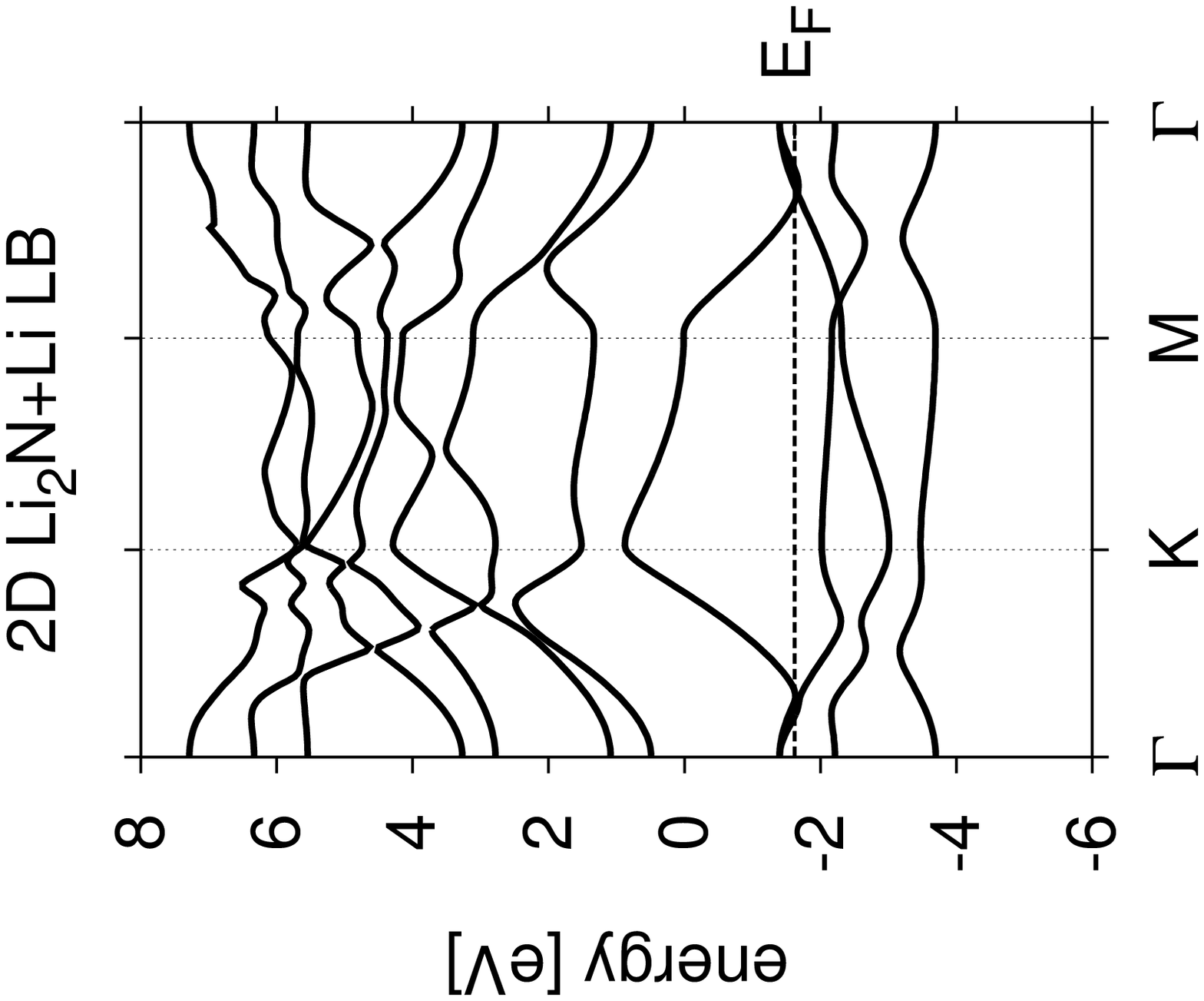}} &
      \resizebox{60mm}{!}{\includegraphics[angle=270, trim=0mm 45mm 0mm 55mm, clip=true]{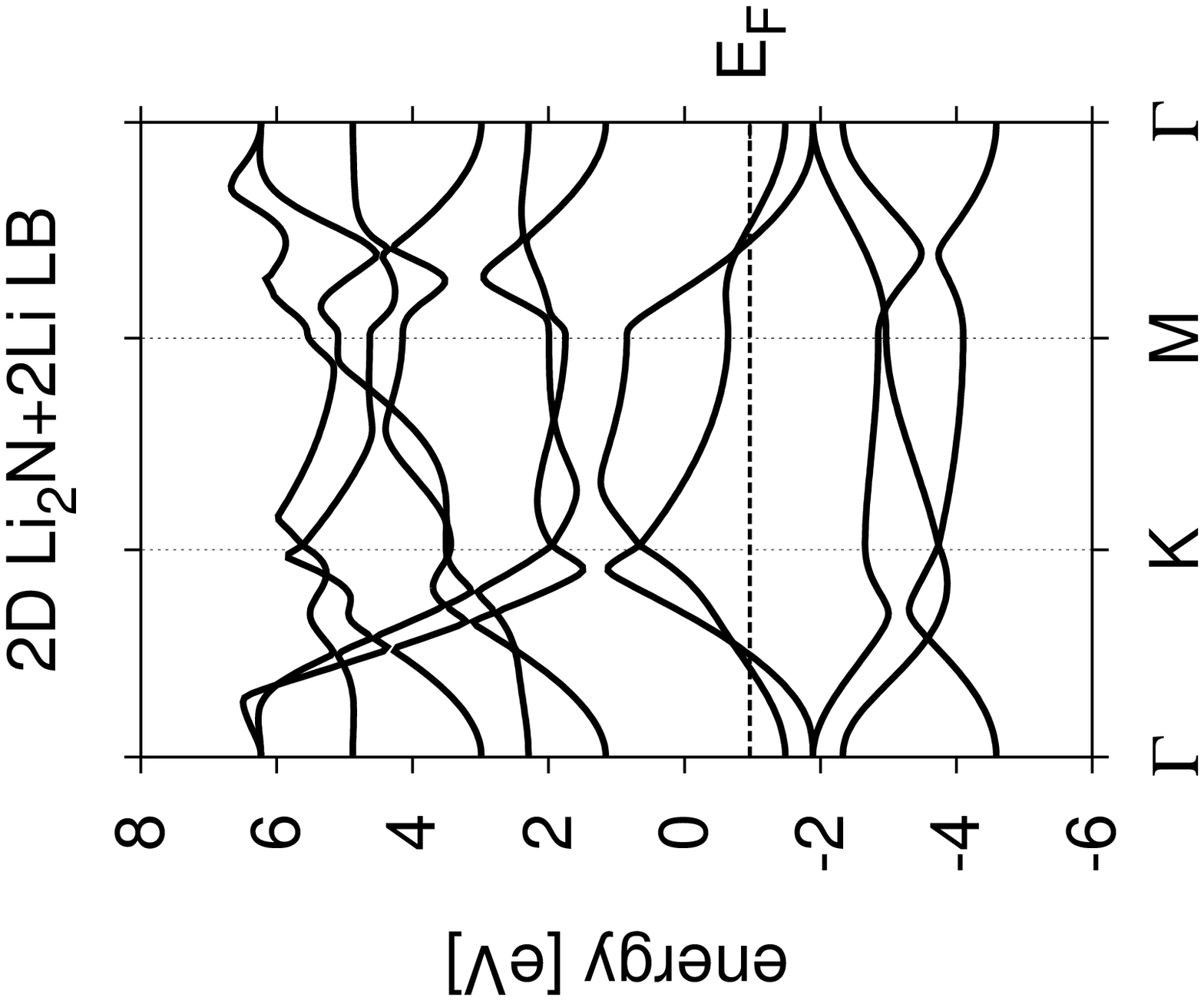}} \\
      \resizebox{60mm}{!}{\includegraphics[angle=270, trim=0mm 45mm 0mm 55mm, clip=true]{pdos_Li2N_LB.ps}} &
			\resizebox{60mm}{!}{\includegraphics[angle=270, trim=0mm 45mm 0mm 55mm, clip=true]{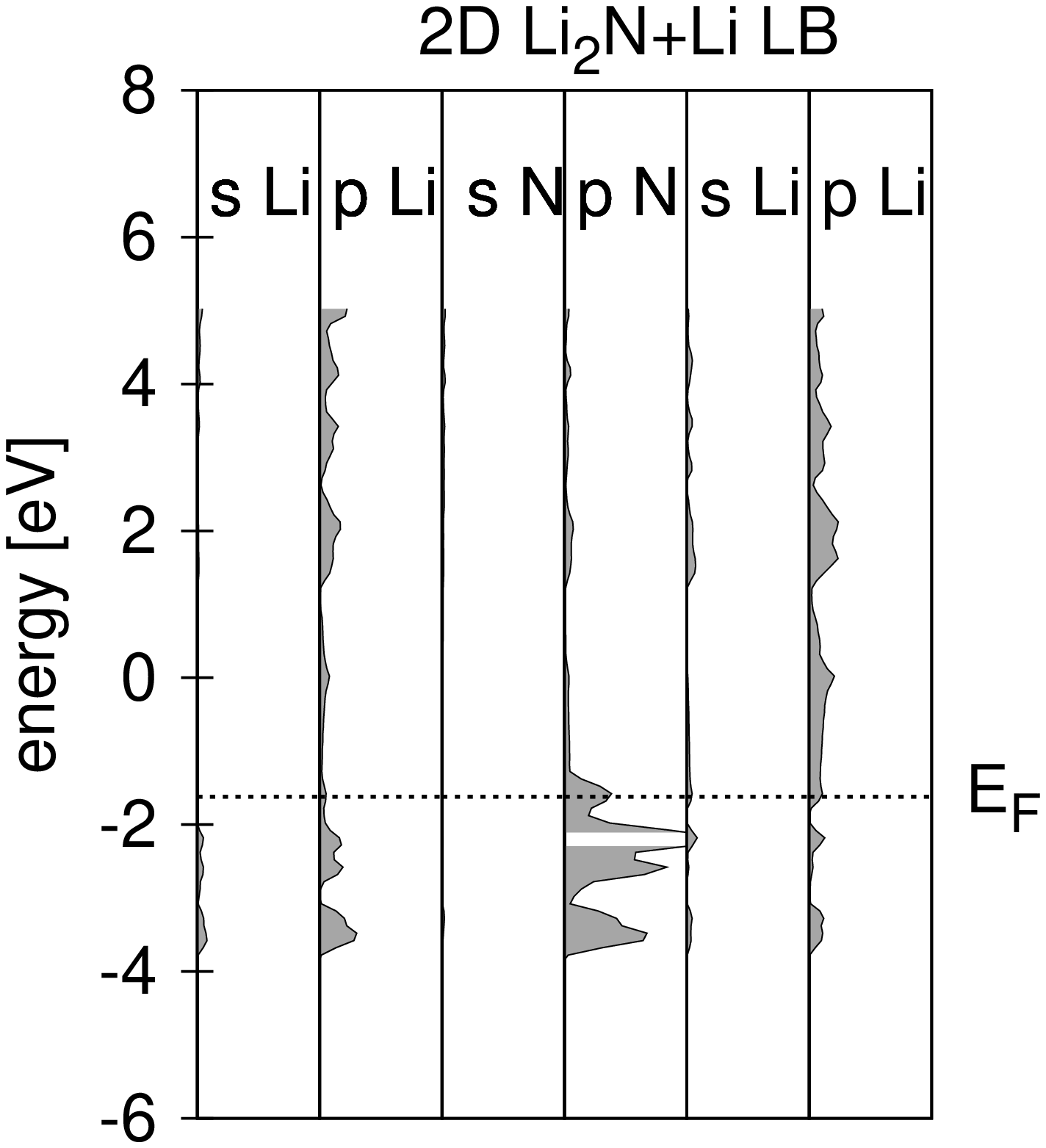}} &
      \resizebox{60mm}{!}{\includegraphics[angle=270, trim=0mm 45mm 0mm 55mm, clip=true]{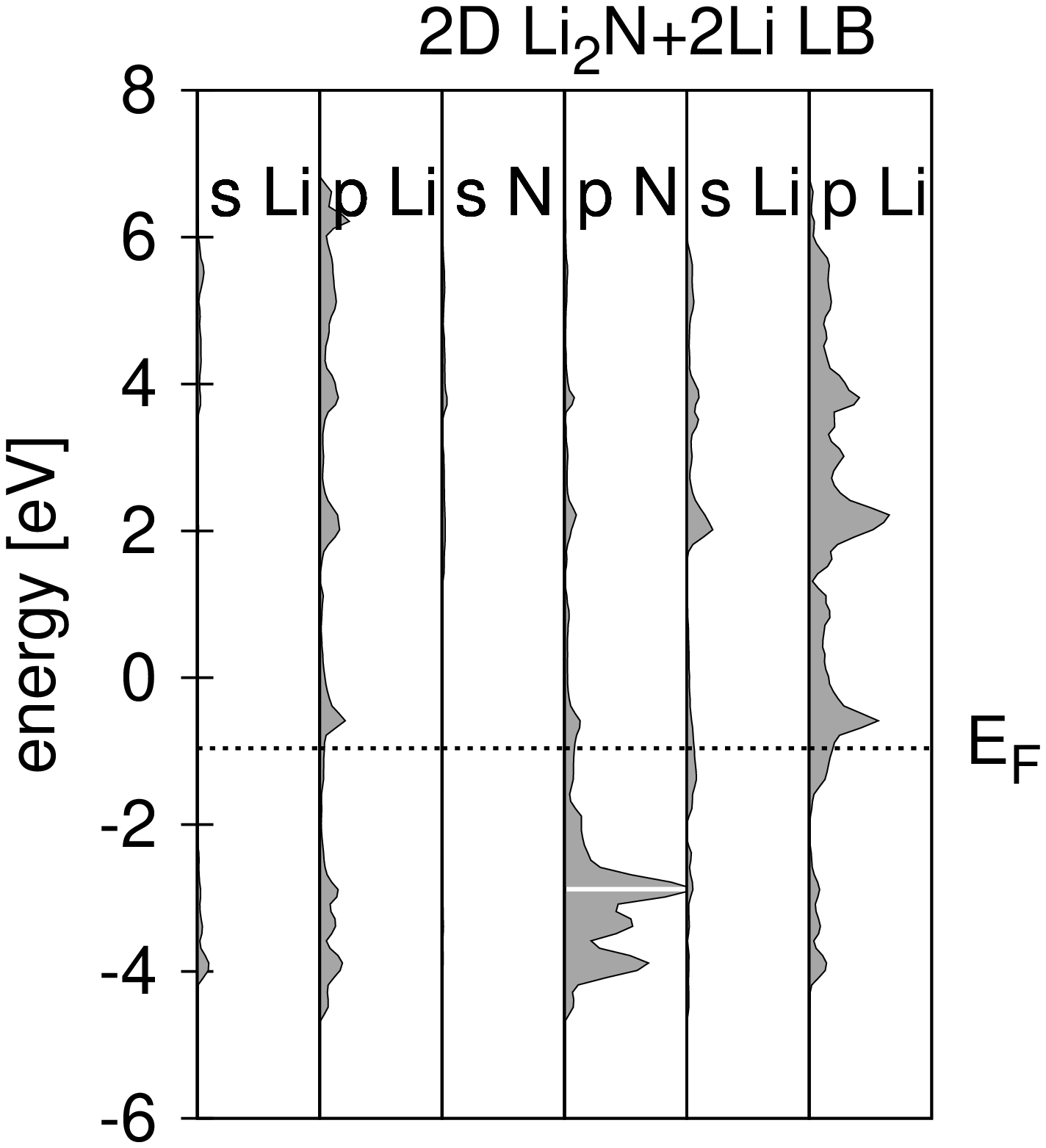}} \\
    \end{tabular}
    \caption{Bandstructures and partial density of states for LB \lin, \linli \ and \linlili. Details in text.}
    \label{fig5}
  \end{center}
\end{figure}


\section{CONCLUSIONS}
$Ab$-$initio$ calculations have been conducted for hypothetical two-dimensional material \lin \ to investigate electronic and magnetic properties.
Calculations show, that structure is much more stable when dangling bonds of nitrogen atoms are functionalized with hydrogen atoms. This hydrogenation
has very strong influence on on bandstructure, changing it from wide-gap semiconductor to metal. \newline
Magnetic properties are also interesting. In analogy to graphene and other two-dimensional materials it is possible to generate non-zero
magnetic moment by introduction of distorsion. In the case of \lin \ the distorsion would be a two-hydrogen or hydrohen-lithium vacancy around the same
nitrogen atom. This generates magnetic moment of 1 $\mu_{B}$. Since bulk Li$_{3}$N material has ususally 1-2\% Li vacancies in Li$_{2}$N layers ~\cite{lin4}
such two-dimensional hydrogenated sheet would be almost naturally magnetic. \newline
These results may give a hint for experimentalists seeking for two-dimensional (magnetic) materials, which would be interesitng addition to
growing family of two-dimensional materials.

\begin{acknowledgments}
Numerical calculations were performed at the Interdisciplinary Centre for Mathematical and Computational Modelling (ICM) at Warsaw University.
\end{acknowledgments}

\end{document}